\documentstyle[eqsecnum,aps]{revtex}

\def\beq{\begin{equation}}
\def\eeq{\end{equation}}

\begin{document}
\draft


\title{A Nearly Minimum Redundant\\
Correlator Interpolation Formula\\
for Gravitational Wave Chirp Detection}


\author{R.P. Croce, Th. Demma}
\address{{\em Wavesgroup, D.I.$^{3}$E., Univ. of Salerno, Italy}}
\author{V. Pierro, I.M. Pinto, F. Postiglione}
\address{{\em Wavesgroup, Univ. of Sannio at Benevento, Italy}\\
}
\date{\today}
\maketitle


\begin{abstract}
An absolute lower bound on the number of templates needed to keep the
fitting factor above a prescribed minimal value $\Gamma$ in correlator-bank
detection of (newtonian) gravitational wave chirps from unknown inspiraling
compact binary stars is derived, resorting to the theory of quasi
band-limited functions in the $L^\infty$ norm. An explicit nearly-minimum
redundant cardinal-interpolation formula for the (reduced, noncoherent)
correlator is introduced. Its computational burden and statistical
properties are compared to those of the plain lattice of (reduced,
noncoherent) correlators, for the same $\Gamma$. Extension to post-newtonian
models is outlined.
\end{abstract}

\pacs{PACS number(s): 04.80.Nn, 95.55.Ym, 95.75.Pq,97.80.Af}



\section*{Introduction}


The detection of gravitational wave (henceforth GW) chirps from unknown
inspiraling compact binary sources (henceforth CBS) is a primary goal for
the early operation of broadband interferometric detectors, including
TAMA300 \cite{TAMA}, GEO600 \cite{GEO}, the two LIGOs \cite{LIGO} and VIRGO 
\cite{VIRGO}, in view of the sizeable expected rate of observable events 
\cite{rate}.

For additive gaussian stationary noise, the correlator-bank
threshold-detector is the optimal one, yielding the smallest false-dismissal
probability, at any fixed false-alarm probability and signal to noise ratio 
\cite{Hels}.

The issues of optimum template parametrization and placement, and the
related computational burden have been discussed by several Authors \cite
{Corr1}-\cite{Hier2}. A lucid account of the main relevant landmarks is
given in \cite{Owe_Sat}.

Curiously, the question of possible {\em efficient} interpolation among the
correlators has been left yet unsolved \cite{Owe_Sat}.

In this paper we set up and test an efficient interpolated representation of
the (reduced, noncoherent) correlator for the simplest paradigm case of
newtonian chirps. The proposed representation \cite{TAMA_pap} is proven to
get close to the {\em absolute minimum} template density required by a
prescribed minimal-match condition, which follows from the theory of
quasi-bandlimited (henceforth q-BL) functions in the $L^\infty$ norm. The
statistical performance of the proposed representation are shown to be
essentially equivalent to those of the plain lattice.

This paper is accordingly organized as follows. In Section I we recall a
number of relevant concepts and results. In Section II we review the design
of the plain template-bank for the simplest (newtonian) case, and discuss
its statistical detection properties. In Section III we briefly introduce
q-BL functions and cardinal expansions, and derive the proposed representations.
In Section IV we compare the computational burden and the statistical
detection/estimation properties of the cardinal-interpolated (reduced,
noncoherent) correlator lattice to those of the plain lattice. Conclusions
follow under Section V.


\section{Background}


In this section we resume a number of well known concepts relevant to CBS
chirp detection, and introduce the notation.


\subsection{Noncoherent Correlator. Deflection and SNR}


Detecting GW chirps from unknown inspiraling CBS requires the computation of
a suitable set of non-coherent correlators (henceforth NCC) \cite{NCC}: 
\begin{equation}
c[\bar{T}]=2\left| \int_{f_{inf}}^{f_{sup}}\frac{A(f)\bar{T}^{*}(f)}{\Pi (f)}%
~df\right| ,  \label{eq:NCC}
\end{equation}
where $(f_{inf},f_{sup})$ is the useful antenna spectral window, $%
A(f)\!=\!S(f)\!+\!N(f)$ are the noise corrupted (spectral) data, resulting
from the superposition of a (possibly null) signal $S(f)$ and a realization $%
N(f)$ of the antenna noise (assumed gaussian and stationary), $\bar{T}(f)$
is an element of a suitable set of unit-norm chirp-{\em templates} such that 
\begin{equation}
||\bar{T}||=\left| 2\int_{f_{inf}}^{f_{sup}}\frac{\bar{T}(f)\bar{T}^{*}(f)}{%
\Pi (f)}~df\right| ^{1/2}\!\!\!=1,  \label{eq:un_norm}
\end{equation}
and $\Pi (f)$ is the (one-sided) antenna noise power spectral density
(henceforth PSD). The random variables $c$ have Ricean probability densities 
\cite{Hels}, 
\begin{equation}
w(c)=c\exp \left( -\frac{c^{2}+d^{2}}{2}\right) I_{0}\left( c~\!d\right) ,
\label{eq:Rice}
\end{equation}
where $I_{0}(\cdot )$ is the modified Bessel function of first kind and zero
order, and 
\begin{equation}
d=\left| 2\int_{f_{inf}}^{f_{sup}}\frac{S(f)\bar{T}^{*}(f)}{\Pi (f)}%
~df\right|  \label{eq:deflec}
\end{equation}
is the {\em deflection} obtained using $\bar{T}$. The moments of (\ref
{eq:Rice}) can be written in terms of Kummer's confluent hypergeometric
function \cite{Rice}: 
\begin{equation}
\langle c^{n}\rangle =2^{n/2}~\Gamma \left( 1+\frac{n}{2}\right)
~_{1}F_{1}\left( -\frac{n}{2};1;-\frac{d^{2}}{2}\right) .  \label{eq:moms}
\end{equation}
For $d\stackrel{>}{\sim }5$ the riceans (\ref{eq:Rice}) merge into
gaussians, with: 
\begin{equation}
E\left[ c\right] \approx d,~~~\mbox{var}\left[ c\right] \approx 1.
\label{eq:moms1}
\end{equation}
The deflection attains its maximum value iff the template $\bar{T}$ is {\em %
matched} to the signal, viz.: 
\begin{equation}
\bar{T}(f)=\frac{S(f)}{||S(f)||},
\end{equation}
yielding: 
\begin{equation}
d=d_{max}=\left| 2\int_{f_{inf}}^{f_{sup}}\frac{S(f)S^{*}(f)}{\Pi (f)}%
~df\right| ^{1/2}=:SNR,
\end{equation}
where SNR is the {\em intrinsic} signal-to-noise ratio.


\subsection{Chirp Templates}


The stationary phase principle (see \cite{StatPhas} for a thorough
discussion of its validity) can be used to show that the asymptotic
principal part \cite{dim_par} of a general \cite{precess}, \cite{eccentr},
reduced \cite{red_mod} post newtonian (henceforth PN) chirp can be written 
\cite{Apo96}: 
\begin{equation}
S(f;\phi _{c},T_{c},\vec{\xi})=Af^{-7/6}\exp \left\{ j\left[ 2\pi
fT_{c}-\phi _{c}+\psi (f,\vec{\xi})\right] \right\} ,  \label{eq:StatPhas}
\end{equation}
where $A$ is a constant (real, unknown) amplitude factor, $T_{c}$ is the
(fiducial) coalescency time \cite{fiducial}, $\phi _{c}$ is the template
phase at $t=T_{c}$, and $\vec{\xi}$ represents the remaining {\em intrinsic}
source parameters \cite{PM_params}.\newline
Equation (\ref{eq:StatPhas}) is used to construct the needed chirp
templates. A further suffix $T$ will be used to label the template
parameters $A_{T}$, $\phi _{c_{T}}$, $T_{c_{T}}$, and $\vec{\xi}_{T}$.%
\newline
All template amplitudes $A_{T}$ will be chosen so as to comply with the
normalization condition (\ref{eq:un_norm}), viz.: 
\begin{equation}
A_{T}=\left[ 2\int_{f_{inf}}^{f_{sup}}\frac{f^{-7/3}}{\Pi (f)}df\right]
^{-1/2}.  \label{eq:un_norm1}
\end{equation}


\subsection{Maximum Likelihood Criterion. Fitting Factor}


Equations (\ref{eq:moms1}) imply that (under the assumption of a uniform
distribution of the unknown source parameters) the {\em largest} correlator
will {\em most likely} correspond to the special template yielding the
largest deflection (maximum likelihood (ML) estimation criterion \cite{Hels}%
).\newline
Data analysis for detecting chirps reduces thus to the following. Given the
(spectral) noisy data, and a set (lattice) of templates, suitably covering
the chirp parameter space, the corresponding (noncoherent) correlators $%
\{c_{k}|k=1,2,\dots ,N\}$ are computed. The {\em largest} among these
correlators is used as a {\em detection statistic} \cite{Hels}, viz.,
whenever this latter exceeds a suitable threshold, set by the prescribed
false-alarm probability ({\em surveillance} strategy \cite{Hels}), a signal
is declared to have been observed \cite{many_sig}, and the corresponding
template is taken as the {\em most likely} estimate of the observed signal 
\cite{Hels}.\newline
It is convenient to measure the {\em goodness of fit} between a given signal 
$S(f)$ and the {\em best available} template in the set in terms of the so
called {\em fitting factor} \cite{FF_credits}: 
\begin{equation}
FF=\max_{k}~\frac{d_{k}}{SNR}.  \label{eq:FF}
\end{equation}
The set of templates should be constructed in such a way that for {\em any}
admissible signal, 
\begin{equation}
FF\geq \Gamma ,  \label{eq:FF_thresh}
\end{equation}
where $1-\Gamma ^{3}$ gauges the fraction of potentially observable sources
which {\em could} be {\em lost} as an effect of template mismatch \cite
{lost_sigs}, \cite{Taylored}.


\subsection{The Reduced Correlator}


In view of eq. (\ref{eq:StatPhas}), the noncoherent correlator (\ref{eq:NCC}%
) can be written: 
\begin{equation}
c=\frac{\displaystyle{\left| 2\int_{f_{inf}}^{f_{sup}}\frac{%
A(f)f^{-7/6}e^{-j\psi _{T}(f,\vec{\xi})]}}{\Pi (f)}\exp (-j2\pi
fT_{c_{T}})df\right| }}{\displaystyle{\left[ 2\int_{f_{inf}}^{f_{sup}}\frac{%
f^{-7/3}}{\Pi (f)}df\right] ^{1/2}}}.  \label{eq:NCC1}
\end{equation}
Equation (\ref{eq:NCC1}) is formally (but for inessential factors) the
absolute value of the ($f\rightarrow T_{c_{T}}$) Fourier transform of the
(complex-valued) function: 
\begin{equation}
K(f)=\left\{ 
\begin{array}{l}
\displaystyle{\frac{A(f)\bar{T}^{*}(f;0,0,\vec{\xi})}{\Pi (f)}}%
,~~~f_{inf}\leq f\leq f_{sup} \\ 
0,~~~~~~~~~~~~~~~~~~~~~~f<f_{inf},~f>f_{sup}.
\end{array}
\right.  \label{eq:kernel}
\end{equation}
Maximizing the noncoherent correlator (\ref{eq:NCC1}) w.r.t. $T_{c_{T}}$, is
thus equivalent to taking the {\em largest} absolute value of the ($%
f\rightarrow T_{c_{T}}$) Fourier transform of (\ref{eq:kernel}). The
resulting {\em reduced} correlator \cite{Ord_Stat} will be denoted with a
capital letter, viz.: 
\begin{equation}
C=\sup_{T_{c_{T}}}~c~.  \label{eq:red_NCC}
\end{equation}


\subsection{The Newtonian Deflection and the Match}


For illustrative purposes, in this paper we shall restrict to the simplest
Newtonian ($0PN$) signals and templates. The $0PN$ function $\psi _{T}(f;%
\vec{\xi})$ in (\ref{eq:StatPhas}) reads: 
\begin{equation}
\psi _{T}(f)=\frac{3}{128}\left( \frac{\pi G}{c^{3}}\right) ^{-5/3}{\cal M}%
_{T}^{-5/3}f^{-5/3},  \label{eq:PN0}
\end{equation}
where ${\cal M}_{T}$ is the template chirp-mass. It is convenient to
introduce the following dimensionless variables and parameters: 
\[
\bar{f}=\frac{f}{f_{inf}},~~\Theta =f_{inf}(T_{c}-T_{c_{T}}),~~\bar{{\cal M}}%
=\frac{{\cal M}}{M_{\odot }}, 
\]
\begin{equation}
\Delta =\bar{{\cal M}}_{s}^{-5/3}-\bar{{\cal M}}_{T}^{-5/3},~~\Lambda =\frac{%
3}{128}\left( \frac{\pi Gf_{inf}M_{\odot }}{c^{3}}\right) ^{-5/3},
\label{eq:ScalVars}
\end{equation}
where $M_{\odot }$ is the solar mass, so as to recast the deflection (\ref
{eq:deflec}) into the form: 
\begin{equation}
d(\Delta ,\Theta )=SNR\cdot \frac{\displaystyle{\left| \int_{1}^{\bar{f}%
_{sup}}d\bar{f}~\frac{\bar{f}^{-7/3}}{\Pi (\bar{f})}\exp \left[ j(2\pi
\Theta \bar{f}+\Lambda \bar{f}^{-5/3}\Delta \right] \right| }}{\displaystyle{%
\int_{1}^{\bar{f}_{sup}}d\bar{f}~\frac{\bar{f}^{-7/3}}{\Pi (\bar{f})}}}.
\label{eq:def_N}
\end{equation}
It is also useful to introduce the {\em reduced} deflection: 
\begin{equation}
D(\Delta )=\max_{\Theta }d(\Delta ,\Theta ),  \label{eq:Phi}
\end{equation}
and the related normalized \cite{bar_fun} functions: 
\begin{equation}
{\bar{d}}(\cdot )=\frac{d(\cdot )}{SNR},~~~{\bar{D}}(\cdot )=\frac{D(\cdot )%
}{SNR}.
\end{equation}
The function ${\bar{D}}(\Delta )$ is known as the (newtonian) {\em match}.
The quantity $\Gamma $ in (\ref{eq:FF_thresh}) is accordingly also named the 
{\em minimal match} \cite{Owe_Sat}.\newline
The functions $\bar{d}$ and $\bar{D}$ are displayed in Fig.s 1 and 2,
respectively, for the special case of a LIGO-like noise PSD, 
\begin{equation}
\Pi (f)=\frac{\Pi _{0}}{5}\left\{ \left( \frac{f_{0}}{f}\right) ^{4}+2\left[
1+\left( \frac{f}{f_{0}}\right) ^{2}\right] \right\} ,  \label{eq:PSD}
\end{equation}
with $f_{0}=300Hz$ ($\Pi _{0}$ is a constant of no concern to us here), and
for a spectral window with: 
\begin{equation}
f_{inf}=40Hz,~~~f_{sup}=400Hz.  \label{eq:window}
\end{equation}
The value $\Theta _{max}$ of $\Theta $ which maximizes $\bar{d}(\Delta
,\Theta )$ in a neighbourhood of $\Delta =0$ is shown in Fig. 3 as a
function of $\Delta $.


\section{The Plain (Newtonian) Lattice}


Given a range $[{\cal M}_{min},{\cal M}_{max}]$ of allowed source chirp
masses, let the set of template chirp-masses be: 
\begin{equation}
\bar{{\cal M}}_{1}^{-5/3}=\bar{{\cal M}}_{max}^{-5/3},~~~\bar{{\cal M}}%
_{n+1}^{-5/3}=\bar{{\cal M}}_{n}^{-5/3}+\delta _{L},~~~n=1,2,\dots ,N_{L}-1,
\end{equation}
where $\delta _{L}$ is the lattice-spacing, and 
\begin{equation}
N_{L}=\left\lceil \frac{\bar{{\cal M}}_{min}^{-5/3}-\bar{{\cal M}}%
_{max}^{-5/3}}{\delta _{L}}\right\rceil  \label{eq:N_Lat}
\end{equation}
is the total number of templates. Obviously $\Delta $ can take only the
discrete values: 
\begin{equation}
\Delta _{n}=\bar{{\cal M}}_{s}^{-5/3}\!-\!\bar{{\cal M}}%
_{n}^{-5/3},~~~n=1,2,\dots ,N_{L}.
\end{equation}


\subsection{Lattice Design}


Given $\bar{{\cal M}}_{s}$, the fitting factor of the lattice is: 
\begin{equation}
FF=\max_{n}\bar{D}(\Delta _{n})=\bar{D}\left( \min_{n}|\Delta _{n}|\right) .
\end{equation}
The minimal-match condition (\ref{eq:FF_thresh}) should be obviously
enforced in the worst case, where: 
\begin{equation}
\bar{{\cal M}}_{s}^{-5/3}=\bar{{\cal M}}_{q}^{-5/3}+\frac{\delta _{L}}{2},
\end{equation}
yielding: 
\begin{equation}
\bar{D}\left( \frac{\delta _{L}}{2}\right) =\Gamma .  \label{eq:spacing}
\end{equation}
Equation (\ref{eq:spacing}) uniquely determines the lattice spacing $\delta
_{L}$, and hence via (\ref{eq:N_Lat}) also the total number of templates.


\subsection{False Alarm and False Dismissal Probabilities}


The statistical distribution of the lattice detection statistic 
\begin{equation}
\max_{k}C_{k}=\max_{h}c_{h}
\end{equation}
is {\em not} known in exact form, and one should resort to numerical
simulations aided by intuition to compute the lattice false-alarm and
false-dismissal probabilities.

The detection threshold $\gamma$ is determined from the prescribed
(tolerated) false alarm probability $\alpha$, by solving the equation: 
\begin{equation}
\alpha = \mbox{prob}[\exists~k : C_k > \gamma| SNR=0] =1-\mbox{prob}%
[\forall~h, c_h < \gamma | SNR=0].  \label{eq:thresh}
\end{equation}
The joint probability in (\ref{eq:thresh}) is difficult to compute, since
the $c_h$ are {\em not} statistically independent in general. For most
practical purposes, a decent approximation is: 
\begin{equation}
\mbox{prob}[\forall~h, c_h < \gamma | SNR=0] = (1-e^{-\gamma^2/2})^M,
\end{equation}
which would be appropriate if the $c_h$ were a collection of $M$ {\em %
independent} ricean random variables. Numerical experiments suggest an
almost linear dependence of $M$ on the total number of NCC used \cite
{Dhu_Schu}.

The probability of {\em false dismissal} of a signal with $SNR \neq 0$ is: 
\begin{equation}
\beta(\gamma,SNR)\!=\!\mbox{prob}[\max_k C_k \!<\! \gamma | SNR\! \neq\! 0]
= \mbox{prob}[\forall h, c_h\!<\! \gamma | SNR \!\neq\! 0].
\label{eq:beta_lat}
\end{equation}
A simple (conservative) approximation of (\ref{eq:beta_lat}) is \cite{Hier1}%
, \cite{false_dism}: 
\begin{equation}
\beta(\gamma,SNR) \approx \mbox{prob}[C_-\!<\!\gamma, C_+\!<\!\gamma | SNR
\neq 0],  \label{eq:beta_bank}
\end{equation}
where $C_-$, $C_+$ denote the reduced correlators corresponding to the
nearest-neighbouring templates, with chirp masses $\bar{{\cal M}}_{\pm}$
such that $\bar{{\cal M}}_s \in [\bar{{\cal M}}_-,\bar{{\cal M}}_+]$. Under
the same (reasonably large SNR) assumptions leading to (\ref{eq:beta_bank}),
the involved joint probability density can be approximated by a gaussian
bivariate \cite{approx}: 
\begin{equation}
w(C_-,C_+) \approx \frac{\displaystyle{\ \exp\left[-\frac{(C_-\! -\!
d_-)^2\!+\!(C_+ \!-\! d_+)^2\!-\!2R(C_- \!-\! d_-)(C_+\!-\!d_+)}{2(1-R^2)}%
\right]}} {2\pi(1\!-\!R^2)^{1/2}},  \label{eq:joint}
\end{equation}
where: 
\begin{equation}
d_{\pm} = d[\Delta_{\pm},\Theta_{max}(\Delta_{\pm})],~~~ R \approx \bar{d}%
[\delta_L,\Theta_{max}(\Delta_+)\!-\!\Theta_{max}(\Delta_-)],
\label{eq:cor_coef}
\end{equation}
and the function $\Theta_{max}(\cdot)$ has been defined in Sect. II.F and
shown in Fig. 3.

Letting: 
\begin{equation}
\bar{{\cal M}}_s^{-5/3}=\bar{{\cal M}}_q^{-5/3} + \eta\delta_L,~~~\eta
\in[0,1[,
\end{equation}
the false dismissal probability (\ref{eq:beta_bank}) is obviously a function
of $\eta$. Within the limits of validity of (\ref{eq:thresh}) to (\ref
{eq:cor_coef}), for a fixed $SNR$ and a prescribed $\alpha$, the following
qualitative dependence of $\beta$ on the lattice spacing $\delta_L$ is
observed. In a neighbourhood of $\eta=0.5$, the false dismissal probability
is reduced by reducing the spacing $\delta_L$ among the templates. On the
other hand, in a neighbourhood of $\eta=0$, reducing the template spacing
produces an {\em increase} of $\beta$. This is due to the dominant effect of
the parallel increase of $\gamma$, needed to keep $\alpha$ unchanged.

A judicious tradeoff should be obviously sought, to choice a value of $%
\Gamma $, and hence of $\delta_L$, via (\ref{eq:spacing}), which minimizes,
e.g., the average value of $\beta$ w.r.t $\eta$, under the assumption of a
uniform distribution of the source chirp mass.


\section{A Nearly Minimum Redundant Interpolated Lattice}


This section contains the main new results. A short introduction to the
theory of q-BL functions is included to make the paper self-contained.


\subsection{q-BL Functions and Cardinal Expansions}


A function $f:x\in [a,b]\rightarrow {\cal R}$ is q-BL in the $L^{\infty }$
norm iff \cite{qBL_th}: 
\begin{equation}
\exists ~\gamma ,B_{c}\in {\cal R}^{+}~:~\sup_{x\in [a,b]}\left|
f(x)-f_{B}(x)\right| =\exp [-\gamma (B-B_{c})],  \label{eq:Linf_err}
\end{equation}
$f_{B}(x)$ being obtained by taking the inverse Fourier transform of the
spectrum of $f(x)$ {\em chopped} at $|y|\geq B$, viz.: 
\begin{equation}
f_{B}(x)={\cal F}_{y\rightarrow x}^{-1}\left\{ W\left( \frac{y}{B}\right)
\cdot {\cal F}_{x\rightarrow y}\left[ f(x)\right] \right\} ,
\label{eq:chop_sp}
\end{equation}
where: 
\begin{equation}
W(x)=\left\{ 
\begin{array}{l}
1,~~~|x|\leq 1 \\ 
0,~~~|x|>1.
\end{array}
\right.  \label{eq:chop}
\end{equation}
For a strictly bandlimited function $f(x)$, whose spectrum vanishes {\em %
identically} outside $[-B,B]$, eq. (\ref{eq:chop_sp}) provides an {\em exact}
interpolating representation known as cardinal expansion \cite{qBL_history}, 
\cite{CS_quick}: 
\begin{equation}
f_{B}(x)=\sum_{n=-\infty }^{\infty }f(x_{n})~\mbox{sinc}\left[ \frac{\pi }{%
\delta }(x-x_{n})\right] ,~~~x_{n+1}\!-\!x_{n}=\delta ,~~~\delta =\frac{1}{2B%
},  \label{eq:cardinal}
\end{equation}
where $\mbox{sinc}(x)=\sin (x)/x$. For a q-BL function on the other hand,
one can prove that eq. (\ref{eq:cardinal}), while reproducing {\em exactly} $%
f(x)$ at $x=x_{k},~k\in {\cal N}$, satisfies eq. (\ref{eq:Linf_err}), i.e.
that 
\begin{equation}
\forall \epsilon >0,~\exists ~B~:~\sup_{x\in [a,b]}\left|
f(x)-f_{B}(x)\right| <\epsilon .  \label{eq:errLinf}
\end{equation}
Equation (\ref{eq:cardinal}) is an {\em approximate} sample-interpolating
representation, where the sample density $\delta ^{-1}=2B$ depends on the
prescribed $L^{\infty }$ approximation error $\epsilon $. It is important to
note that the {\em exponential} decay of the error in (\ref{eq:Linf_err})
implies that reducing $\epsilon $ in (\ref{eq:errLinf}) by orders of
magnitude does {\em not} change the order of magnitude of $B$.\newline

Usually one needs to compute $f(x)$ in a {\em finite} interval $[a,b]$
including only 
\begin{equation}
N =\left\lceil \frac{(b-a)}{\delta} \right\rceil  \label{eq:NShannon}
\end{equation}
samples \cite{degs_freed}. However, using eq. (\ref{eq:cardinal}) to compute 
$f(x)$ in $[a,b]$ requires, in principle, knowledge of an {\em infinite}
number of samples {\em outside} the interval of interest. This limitation
can be circumvented by using generalized ({\em economized}) cardinal
expansions. These expansions have the general form \cite{Knab}: 
\[
f(x)= \sum_{n=-\infty}^{\infty} f(x_n)~\mbox{sinc} \left[\frac{\pi}{%
\delta^{\prime}}(x-x_n)\right]~ \theta(x-x_n), 
\]
\begin{equation}
x_{n+1}-x_n = \delta^{\prime},~~~ \delta^{\prime}=(2 \chi B)^{-1},~~ \chi > 1
\label{eq:smart}
\end{equation}
where $\theta(x)$ is a suitable {\em windowing function} such that: 
\begin{equation}
\theta(0)=1,~~~{\cal F}_{x \rightarrow \xi} \left[\theta(x)\right] =
0,~\forall~\xi > (\chi-1)B,~~~\chi > 1.  \label{eq:constraints}
\end{equation}
The expansion (\ref{eq:smart}) is nothing but the$~$std. cardinal expansion
of the function $f(x)\theta(u-x)$, whose bandwidth under the assumptions (%
\ref{eq:constraints}) is $B^{\prime}=\chi B$, viz.: 
\[
f(x)\theta(u-x)=\sum_{n=-\infty}^{\infty} f(x_n)\theta(u-x_n)~ \mbox{sinc}%
\left[\frac{\pi}{\delta^{\prime}}\left(x-x_n\right)\right], 
\]
\begin{equation}
x_{n+1}-x_n=\delta^{\prime},~~~ \delta^{\prime}=\frac{1}{2 \chi B},
\label{eq:smooth1}
\end{equation}
evaluated at $u=x$. The Fourier spectrum of $f(x)\theta(u-x)$ is a {\em %
smoothed} version of the plain spectrum of $f(x)$; a judicious choice of $%
\theta(x)$ can thus make in principle, the {\em decay rate} of $%
f(x_n)\theta(u-x_n)$ as $|u-x_n| \rightarrow \infty$ as fast as desired \cite
{smooth}.\newline
The Knab window function \cite{Knab}: 
\begin{equation}
\theta(x)=K_P(x):= \frac{\mbox{sinh} \left\{ \pi P (1\!-\!\chi^{-1}) \left[
1-\displaystyle{\left(\frac{x}{P\delta^{\prime}}\right)}^2 \right]^{1/2}
\right\} } {\left[ 1-\displaystyle{\left(\frac{x}{P\delta^{\prime}}\right)}%
^2 \right]^{1/2} \mbox{sinh} [\pi P (1\!-\!\chi^{-1}) ] }  \label{eq:Knab}
\end{equation}
satisfies all constraints (\ref{eq:constraints}) and is {\em essentially}
confined \cite{Knab_conf} in $|x| \leq P\delta^{\prime}$. This allows to
truncate (\ref{eq:smart}) at $|x-x_n| \approx P\delta^{\prime}$, so that for
any given $x$, {\em only} $\approx 2P$ samples symmetrically placed around $%
x $ are {\em essentially} needed to reconstruct $f(x)$.\newline
The error resulting from truncation of (\ref{eq:smart}) with (\ref{eq:Knab})
at $|x-x_n| = P\delta^{\prime}$ has been discussed in \cite{Knab}. A simple
(and conservative) upper bound is given by: 
\[
\left| \sum_{|x-x_n|>P\delta^{\prime}} f(x_n)~\mbox{sinc} \left[\frac{\pi}{%
\delta^{\prime}}(x-x_n)\right]~ K_P(x-x_n) \right| < 
\]
\begin{equation}
< \frac{M}{\mbox{sinh}[\pi P(1-\chi^{-1})]},~~~ M=\sup_{x\in[a,b]}~f(x).
\label{eq:KN_err}
\end{equation}
Usually, one enforces the condition: 
\begin{equation}
\frac{M}{\mbox{sinh}[\pi P(1-\chi^{-1})]} = \epsilon^{\prime}\ll \epsilon,
\label{eq:bnd1}
\end{equation}
where $\epsilon$ is the prescribed $L^\infty$ error in (\ref{eq:errLinf}).
Equation (\ref{eq:bnd1}) can be solved to express $P$ as a function of $\chi$%
, 
\begin{equation}
P=\frac{\mbox{sinh}^{-1}\left(M/\epsilon^{\prime}\right) }{ \pi(1-\chi^{-1}) 
}.  \label{eq:P_vs_chi}
\end{equation}
The total number of samples 
\begin{equation}
N_T= \left\lceil \frac{(b-a)}{\delta^{\prime}} \right\rceil+2P= \left\lceil
\chi \frac{(b-a)}{\delta} +2P \right\rceil  \label{eq:Nsamp}
\end{equation}
needed to represent $f(x)$ in $[a,b]$ using (\ref{eq:smooth1}) and (\ref
{eq:Knab}), within a prescribed $L^\infty$ error and under the constraint (%
\ref{eq:bnd1}), can be accordingly minimized by letting: 
\begin{equation}
\chi = 1+ \left[ \frac{2\delta~\mbox{sinh}^{-1}(M/\epsilon^{\prime})}{\pi
(b-a)} \right]^{1/2}.  \label{eq:chi_opt}
\end{equation}
\[
~ 
\]


\subsection{The Cardinal-Interpolated Newtonian Match}


The match $\bar{D}(\Delta )$ is a q-BL function in the $L^{\infty }$ norm.
This can be seen from Fig. 4, where the {\em exponential} decay of the $%
L^{\infty }$ error in $[0,\infty [$ between $\bar{D}(\Delta )$ and the
cardinal expansion: 
\begin{equation}
\bar{D}_{B}(\Delta )=\sum_{n=-\infty }^{\infty }\bar{D}(\Delta _{n})%
\mbox{sinc}\left[ \frac{\pi }{\delta _{C}}\left( \Delta \!-\!\Delta
_{n}\right) \right] ,~~~\Delta _{n+1}-\Delta _{n}=\delta _{C},  \label{eq:DB}
\end{equation}
is displayed as a function of $\delta _{C}^{-1}$ on a Log-Lin plot \cite
{also_L2}. Switching back to the original variables, eq. (\ref{eq:DB})
reads: 
\begin{equation}
\bar{D}_{B}(\bar{{\cal M}}_{s}^{-5/3}\!-\!\bar{{\cal M}}_{T}^{-5/3}):=\!\!%
\sum_{n=-\infty }^{\infty }\!\!\bar{D}(\bar{{\cal M}}_{s}^{-5/3}\!\!-\!\bar{%
{\cal M}}_{n}^{-5/3})\cdot \mbox{sinc}\left[ \frac{\pi }{\delta _{C}}%
\!\left( \bar{{\cal M}}_{T}^{-5/3}\!\!-\!\bar{{\cal M}}_{n}^{-5/3}\right)
\right] ,  \label{eq:ESID}
\end{equation}
where \cite{neg_mass}: 
\begin{equation}
\bar{{\cal M}}_{n+1}^{-5/3}-\bar{{\cal M}}_{n}^{-5/3}=\delta _{C}.
\end{equation}
Given $\bar{{\cal M}}_{s}$, the fitting factor obtained using (\ref{eq:ESID}%
) is given by: 
\begin{equation}
FF=\max_{\bar{{\cal M}}_{T}}\bar{D}_{B}(\bar{{\cal M}}_{s}^{-5/3}-\bar{{\cal %
M}}_{T}^{-5/3})=:\bar{D}_{B}(\bar{{\cal M}}_{s}^{-5/3}-\bar{{\cal M}}%
_{*}^{-5/3}).  \label{eq:FFdef}
\end{equation}
It is convenient to let: 
\begin{equation}
\bar{{\cal M}}_{s}^{-5/3}=\bar{{\cal M}}_{q}^{-5/3}+\eta \delta _{C},~~~\eta
\in [0,1[,
\end{equation}
\begin{equation}
\bar{{\cal M}}_{*}^{-5/3}=\bar{{\cal M}}_{q}^{-5/3}+\eta _{*}\delta
_{C},~~~\eta _{*}\in [0,1[,
\end{equation}
so as to rewrite (\ref{eq:FFdef}) as: 
\begin{equation}
FF=\bar{D}_{B}(\eta -\eta _{*}).  \label{eq:FFa}
\end{equation}
The difference $\eta -\eta _{*}$ turns out to depend on $\eta $ as shown in
Fig. 5, and hence the fitting factor (\ref{eq:FFa}) depends on $\eta $ as
shown in Fig. 6. The minimal-match condition (\ref{eq:FF_thresh}) should
again be enforced in the worst case(s), i.e., as seen from Fig.s 5, 6, for $%
\eta =0.5$ and $\eta _{*}=\eta $, yielding: 
\begin{equation}
\sum_{n=-\infty }^{\infty }\bar{D}\left[ \left( n+\frac{1}{2}\right) \delta
_{C}\right] \cdot \mbox{sinc}\left[ \left( n+\frac{1}{2}\right) \pi \right]
=\Gamma .  \label{eq:cond}
\end{equation}
This condition is notably {\em indipendent} on $q$, and fixes the sample
spacing $\delta _{C}$.\newline

In practice, as discussed in the previous section, it is convenient to use
an {\em economized} cardinal expansion, viz: 
\begin{equation}
\bar{D}_B(\bar{{\cal M}}_s^{-5/3}\!\!-\!\!\bar{{\cal M}}_T^{-5/3})\! :=\!\!
\sum_{n=-\infty}^{\infty} \bar{D}(\bar{{\cal M}}_s^{-5/3}\!\!-\!\!\bar{{\cal %
M}}_n^{-5/3}) \Psi_n \left( \bar{{\cal M}}_T^{-5/3}\!\!-\!\!\bar{{\cal M}}%
_n^{-5/3} \right),  \label{eq:ESID1}
\end{equation}
where (see eq.s (\ref{eq:smart}) and (\ref{eq:Knab})): 
\begin{equation}
\bar{{\cal M}}_{n+1}^{-5/3}-\bar{{\cal M}}_n^{-5/3}=\delta_C^{\prime}=%
\chi^{-1}\delta_C,  \label{eq:chmass_smp}
\end{equation}
and: 
\begin{equation}
\Psi_{n}(x)=\mbox{sinc} \left( \frac{\pi x}{\delta_C^{\prime}} \right) \cdot 
\frac{\mbox{sinh} \left\{ \pi P (1\!-\!\chi^{-1}) \left[ 1-\displaystyle{%
\left(\frac{x}{P\delta_C^{\prime}}\right)}^2 \right]^{1/2} \right\} } {%
\left[ 1-\displaystyle{\left(\frac{x}{P\delta_C^{\prime}}\right)}^2
\right]^{1/2} \mbox{sinh} [\pi P (1\!-\!\chi^{-1}) ] }.  \label{eq:Interp}
\end{equation}
As shown in the previous section, the function (\ref{eq:Interp}) is {\em %
essentially} contained in the interval: $|x| < P\delta_C^{\prime}$, and
hence the infinite sum in (\ref{eq:ESID1}) is {\em essentially} restricted
to: 
\begin{equation}
\left\lceil \frac{\bar{{\cal M}}_T^{-5/3}}{\delta_C^{\prime}} \right\rceil -
P \leq n \leq \left\lfloor \frac{\bar{{\cal M}}_T^{-5/3}}{\delta_C^{\prime}}%
\right\rfloor + P.  \label{eq:trunc}
\end{equation}
Capitalizing on eq. (\ref{eq:KN_err}) we shall enforce the condition: 
\begin{equation}
\frac{1}{\mbox{sinh}[\pi P(1-\chi^{-1})]} = \frac{1-\Gamma}{10}
\label{eq:tr_err}
\end{equation}
to guarantee that the minimal-match condition will not be affected within
the last significant figure of $\Gamma$, when using using eq. (\ref{eq:ESID1}%
) truncated according to (\ref{eq:trunc}) in place of eq. (\ref{eq:ESID}).
Further, in view of eq.s (\ref{eq:P_vs_chi}) and (\ref{eq:chi_opt}), we
shall take: 
\begin{equation}
\chi=\chi_{opt}=: 1+ \left[ \frac{2\delta_C~\mbox{sinh}^{-1} \displaystyle{%
\left(\frac{10}{1-\Gamma}\right)} }{ \pi \left( \bar{{\cal M}}_{min}^{-5/3}-%
\bar{{\cal M}}_{max}^{-5/3} \right) } \right]^{1/2}  \label{eq:chiopt}
\end{equation}
and: \vspace*{-3mm} 
\begin{equation}
P=P_{opt}=: \frac{\mbox{sinh}^{-1} \displaystyle{\left(\frac{10}{1-\Gamma}%
\right)} }{ \pi(1-\chi_{opt}^{-1}) },  \label{eq:Popt}
\end{equation}
\[
~ 
\]
so as to minimize the total number of correlators 
\begin{equation}
N_C = \left\lceil \chi~\frac{\bar{{\cal M}}_{min}^{-5/3}-\bar{{\cal M}}%
_{max}^{-5/3}}{\delta_C} +2P \right\rceil.
\end{equation}
needed to evaluate (\ref{eq:ESID1}) throughout the range $[{\cal M}_{min}, 
{\cal M}_{max}]$ of ${\cal M}_T$.


\subsection{The Cardinal-Interpolated Reduced Correlator}


As a next step, we make the {\em ansatz} that an approximate representation
of the reduced (newtonian) noncoherent correlator in terms of a
(generalized) cardinal expansion also holds, viz. \cite{stoch_CS}: 
\begin{equation}
C_{B}:=\sum_{n=-\infty }^{\infty }C_{n}\Psi _{n}(\bar{{\cal M}}_{T}^{-5/3}-%
\bar{{\cal M}}_{n}^{-5/3}),  \label{eq:C_NSK}
\end{equation}
where: 
\begin{equation}
C_{n}=\max_{T_{c_{T}}}\left| 2\int_{f_{inf}}^{f_{sup}}\frac{A(f)\bar{T}%
_{n}^{*}(f)}{\Pi (f)}df\right| ,  \label{eq:main}
\end{equation}
and the infinite sum is truncated according to (\ref{eq:trunc}).\newline
In (\ref{eq:main}) the templates $\bar{T}_{n}$ are defined by eq.s (\ref
{eq:StatPhas}), (\ref{eq:un_norm1}) and (\ref{eq:PN0}), where the (scaled)
chirp masses take the values 
\begin{equation}
\bar{{\cal M}}_{k}^{-5/3}=\bar{{\cal M}}_{max}^{-5/3}+k\delta _{C}^{\prime
},~~~k=-P,-P+1,\dots ,N_{C}-1+P,  \label{eq:tmplts}
\end{equation}
the interpolating functions $\Psi _{n}(x)$ are given by (\ref{eq:Interp}),
and the parameters $\delta _{C}$, $\chi $ and $P$ are computed from the
prescribed minimal match $\Gamma $ as explained In Sect. III.A.


\section{Plain vs. Cardinal Interpolated Lattice}


In this section we shall compare the (newtonian) cardinal-interpolated
(reduced, noncoherent) correlator to the plain-lattice of (reduced
noncoherent) correlators in terms of computational cost and statistical
features (detection and estimation performance). The assumed noise PSD and
spectral window are given by (\ref{eq:PSD}) and (\ref{eq:window}),
respectively.


\subsection{Computational Burden}

The plain lattice template spacings $\delta _{L}$ and total number of
correlators $N_{L}$ needed to cover the range $(0.2M_{\odot },10M_{\odot })$
for some values of the minimal match $\Gamma $ are compared in Table I to
the corresponding quantities $\delta _{C}$ and $N_{C}$ of the cardinal
interpolated correlator.

It is seen that at {\em any} value of the minimal match $\Gamma$, the
cardinal-interpolated representation requires some $30\%$ {\em less many}
templates than the plain lattice.

On the other hand, evaluating (\ref{eq:C_NSK}) at any value ${\cal M}_T \neq 
{\cal M}_k$ is substantially cheaper than computing the corresponding
(reduced, noncoherent) correlator.

Indeed, to use (\ref{eq:C_NSK}) one really needs to evaluate the
interpolating functions $\Psi_n(x)$ only at a finite number of (equispaced)
values of $\bar{{\cal M}}_T^{-5/3}$ between the samples $\bar{{\cal M}}%
_q^{-5/3}$. The corresponding values of the interpolating functions can be
computed once for all, and stored in a look-up table. As a result, only $%
\sim 2P$ floating point operations are needed to compute (\ref{eq:C_NSK}) at
each of the above values of $\bar{{\cal M}}_T^{-5/3}$, with a typical $P
\sim 10^2$. 

\subsection{Statistical Features}


The statistical properties of the cardinal-interpolated correlator have been
compared to those of the plain-lattice via extensive Monte Carlo
simulations. The number of different realizations used to derive the
statistics was $\sim 10^{4}$.

Simulated data were sampled in time at twice the Nyquist rate. To limit
running times, the minimal-match was set at $\Gamma=0.9$, and the chirp-mass
range was chosen in such a way that the longest observable waveform spanned $%
2^{15}$ time bins. In order to avoid circular-correlation artifacts, and to
have equal statistics for all reduced correlators, all templates were
zero-padded up to a total length of $2^{16}$ bins. Gaussian uncorrelated
noise samples were generated using a feedback-shift-register routine from
the IMSL package, featuring an extremely large period \cite{RAN}, followed
by a Box-M\"{u}ller transformation \cite{NumRec}. The noise samples were
added to the whitened data in the spectral domain.

In the following we shall denote the cardinal-interpolated and plain lattice
test-statistics as \cite{Max_What}: 
\begin{equation}
C^{(C)}_{max}= \max_{{\cal {M}_T}} \sum_{k} C_k \Psi_k \left( \bar{{\cal M}}%
_T\!\!-\!\!\bar{{\cal M}}_{k} \right)  \label{eq:Cmax_Card}
\end{equation}
and: 
\begin{equation}
C^{(L)}_{max} = \max_{k} C_k.  \label{eq:Cmax_Lat}
\end{equation}
respectively. We shall denote the estimated mass, i.e., the value of $\bar{%
{\cal M}}_T$ which yields the maximum in (\ref{eq:Cmax_Card}) or (\ref
{eq:Cmax_Lat}) as ${\cal M}_{est}$, and let: 
\begin{equation}
\bar{{\cal M}}_{est}^{-5/3}\!\!=\!\bar{{\cal M}}_q^{-5/3}\!\!+\!\eta_{est}%
\delta,~~~ \eta_{est} \in [0,1[,  \label{eq:est_M}
\end{equation}
\begin{equation}
\bar{{\cal M}}_s^{-5/3}\!\!=\!\bar{{\cal M}}_q^{-5/3}\!\!+\!\eta\delta,~~~
\eta \in [0,1[,  \label{eq:source_M}
\end{equation}
Whenever needed a suffix/superfix $L,C$ will be used to identify the
plain-lattice and cardinal-interpolated cases in (\ref{eq:est_M}), (\ref
{eq:source_M}).

The CDFs of $C^{(C)}_{max}$ (dashed lines) and $C^{(L)}_{max}$ (full lines)
in the absence of signal ($SNR\!=\!0$), are compared in Fig. 7-a, for
template spacings corresponding to $\Gamma=0.9$. The corresponding PDFs are
displayed in Fig. 7-b. The observed difference falls within the $3\sigma$
uncertainty interval related to the finite number of realizations.

The CDFs of $C^{(C)}_{max}$ (dashed lines) and $C^{(L)}_{max}$ (full lines)
in the presence of a signal with $SNR\!=\!6,8,10$ are shown in Fig. 8-a, for
the (worst) case where $\eta=0.5$ in (\ref{eq:source_M}). The corresponding
PDFs are displayed in Fig. 8-b. Again, the observed differences fall within
the $3\sigma$ uncertainty interval related to the finite number of
realizations. Note that the expected value always exceeds the design value $%
\Gamma \cdot SNR$, as might be expected as an effect of the supremum-taking
operations in (\ref{eq:red_NCC}), (\ref{eq:Cmax_Card}) and (\ref{eq:Cmax_Lat}%
).

As $\eta$ in (\ref{eq:source_M}) changes between $0$ and $0.5$, the PDFs of $%
C^{(C)}_{max}$ and $C^{(L)}_{max}$ change in turn. The limiting PDFs
corresponding to $\eta\!=\!0$ and $\eta\!=\!0.5$ are shown in Fig.s 9 and
10, for the special case $SNR\!=\!8$, for of $C^{(C)}_{max}$ and $%
C^{(L)}_{max}$.

It can be concluded that the detection performance of the cardinal
interpolated (reduced, noncoherent) correlator is essentially equivalent to
that of the computationally more expensive plain lattice of (reduced,
noncoherent) correlators.

We turn now to a comparison of the pertinent {\em estimation} features. To
this end we let: 
\begin{equation}
\xi= \frac{\bar{{\cal M}}_{est}^{-5/3}-\bar{{\cal M}}_{s}^{-5/3}} {\delta}=
\eta-\eta_{est}.
\end{equation}

The PDFs of $\xi$ for the cardinal-interpolated correlator at $SNR=8$ and $%
\Gamma=0.9$ is shown in Fig. 11 for $\eta=0$ and $\eta=0.5$.

The corresponding probabilities $P(\xi)$ for the plain lattice of
correlators are shown in Fig. 12. Note that for the plain lattice of
correlators, $\eta_{est}$ can take only values which are integer multiples
of $\delta_L$.

Both the cardinal-interpolated correlator and the plain lattice of
correlators provide {\em biased} estimates. The bias $E[\xi]$ becomes for
both essentially independent of the $SNR$ at sufficiently high $SNR$ levels (%
$SNR \stackrel{>}{\sim} 8$). For the cardinal-interpolated correlator the
asymptotic large-SNR bias is shown in Fig. 13, which is a close akin of Fig.
5. For the plain lattice, it is displayed in Fig. 14. The
cardinal-interpolated correlator is seen to exhibit a smaller bias.

The std. deviations of the cardinal-interpolated and plain lattice
estimators are nearly the same. For instance, for $\eta=0.5$ (worst case),
at $SNR=8$ one has $\sigma(\xi_L)=1.12$ and $\sigma(\xi_C)=.92$; at $SNR=10$ 
$\sigma(\xi_L)=0.84$ and $\sigma(\xi_C)=0.63$.


\subsection{Extension to PN Models}


The cardinal-interpolated approach can be extended in principle, to higher
order PN models, provided the {\em structure} and q-BL properties of the
(reduced) correlator are preserved. In the geometrical language of \cite
{Owen1}, this is equivalent to requiring that the chosen parameter space be
(globally) flat and Euclidean \cite{PN_sampling}. This is surely the case
for 1PN models \cite{Owe_Sat}, and almost the case for the new $0-$spin 2PN
coordinates proposed in \cite{Tagoshi_Tanaka}. One should expect an even
more substantial computational saving, in view of the higher dimension of
the parameter space.


\section{Conclusions and Recommendations}


Quasi-bandlimited function approximation theory can be used to build a
(nearly) minimum redundant cardinal-interpolated representations of the
noncoherent correlator for detecting gravitational wave chirps. An explicit
expression has been provided and tested, for the simplest case of newtonian
waveforms.

The number of correlators to be computed and interpolated in order to
maintain the match above a given minimal value $\Gamma$ has been shown to be 
{\em substantially less} then required by the std. (lattice) approach, and
the computational gain goes up with $\Gamma$. On the other hand, evaluating
the cardinal-interpolated representation at any value of ${\cal M}_T$ is
substantially cheaper than computing the corresponding correlator.

We suggest that cardinal-interpolated expansions could be used to improve
the efficiency of hierarchical searches, at all hierarchical levels.

Extension to PN templates should be straighforward, in principle, insofar as
the {\em structure} and q-BL property of the correlators is preserved, and
lead to an even increased computational gain. Work in this direction is in
progress.


\section{Acknowledgements}


This work has been sponsored in part by the European Community through a
1998 Senior Visiting Scientist Grant to I.M. Pinto at NAO - Spacetime
Astronomy Division, Tokyo, Japan, in connection with the TAMA project. I.M.
Pinto wishes to thank all the TAMA staff at NAO, and in particular prof.
Fujimoto Masa-Katsu and prof. Kawamura Seiji for kindly hospitality and
stimulating discussions.





\begin{references}
\bibitem{fiducial}  {$T_c$ is the time where the model's instantaneous
orbital period goes to zero. The inspiral-model however ceases to be valid
well before.}
\bibitem{bar_fun}  {The functions $d$ and $D$ attain their maximum value iff
the template is perfectly matched to the signal, viz. $\Delta\!=\!\Theta\!=%
\!0$, yielding $d\!=\!D\!=\!SNR$ and $\bar{d}\!=\!\bar{D}\!=\!1$.}
\bibitem{CS_quick}  {The cardinal expansion (\ref{eq:cardinal}) is most
simply obtained by regarding the inverse Fourier transform representation: 
\[
f(x)= \int_{-B}^{B}dy~ F(y)\exp(j2\pi x y) 
\]
as a $L^2_{[-B,B]}$ scalar product, and using Parseval identity, viz.: 
\[
\langle f,g \rangle = \sum_n \langle f,u_n \rangle \langle g,u_n \rangle^* 
\]
where $f(y)\!=\!F(y)$, $g(y)\!=\!\exp(-j2\pi x y)$, and $u_n(y)\!=%
\!(2B)^{-1/2}\exp(jn\pi y/B)$ is the Euler-Fourier basis on $[-B,B]$.}
\bibitem{Knab_conf}  {More specifically, its $L^\infty$ (and $L^2$) norm in 
$|x| > Q\delta^{\prime}$ vanishes exponentially in $(Q\!-\!P)$. Knab's
window (\ref{eq:Knab}) is a slightly sub-optimal choice for $\theta(x)$
under the constraints (\ref{eq:constraints}). It provides possibly the best
available tradeoff between the competing requirements of minimal
oversampling and minimal computational burden \cite{Knab}. Alternative
choices are discussed in \cite{Various_Win}.}
\bibitem{also_L2}  {It can be shown that $\bar{D}(\Delta)$ is a q-BL
function in the $L^2$ norm as well.}.
\bibitem{stoch_CS}  {The extension of (generalized) cardinal expansions to
stochastic processes is due to Kotel'nikov, and is discussed e.g., in \cite
{Various_Win}. Note in this connection that the additive noise affecting the
data does not depend on ${\cal M}$.}
\bibitem{Max_What}  {The maximum in (\ref{eq:Cmax_Lat}) is actually taken on
a discrete set of values of $\bar{{\cal M}}_T^{-5/3}$ which contains $10$
(equispaced) samples in each interval $[\bar{{\cal M}}_q^{-5/3}, \bar{{\cal M%
}}_q^{-5/3}+\delta_C^{\prime}]$.}
\bibitem{PN_sampling}  {Under these assumptions, a straightforward
multi-dimensional cardinal expansion can be written, for which the sampling
intervals $\delta^i_C$ are uniform and mutually independent.}
\end{references}
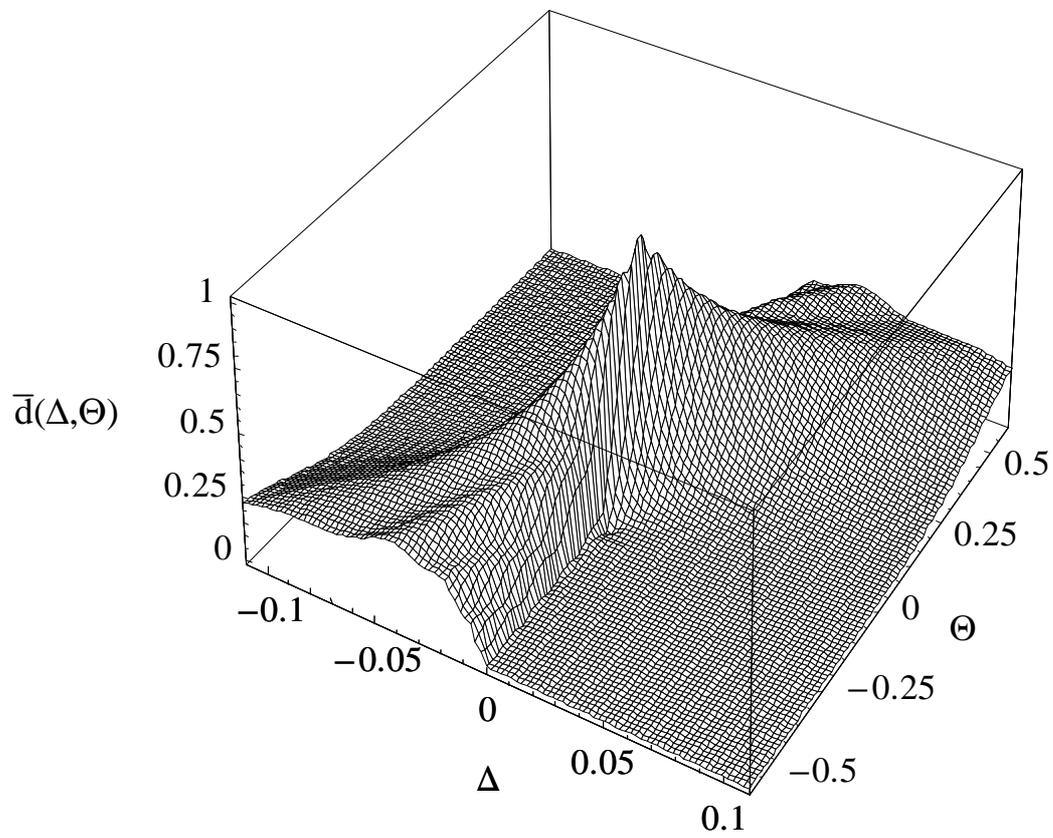
\end{document}